\documentclass[12pt]{article}
\usepackage{authblk}
\usepackage{mathptmx}       
\usepackage{helvet}         
\usepackage{courier}        
\usepackage{type1cm}        
\usepackage{atbegshi}
\AtBeginDocument{\AtBeginShipoutNext{\AtBeginShipoutDiscard}}

\usepackage{graphicx}        

\def \ee {\begin{equation}}
\def \eee {\end{equation}}
\def \eqe {\begin{eqnarray}}
\def \eqee {\end{eqnarray}}

\begin{document}
\title{Excitable behaviors}
\thanks{RS is with the Department of Engineering, University of Cambridge, United Kingdom. Email: rs771@cam.ac.uk.  GD is with the Institut Montefiore, Universit\'e de Liege, Belgium.  Email: g.drion@ulg.ac.be. 
AF is with Department of Mathematics,  Universidad Nacional Aut\'onoma de M\'exico, Mexico. Email: afranci@ciencias.unam.mx.The research leading to these results has received funding from the European Research Council under the Advanced ERC Grant Agreement Switchlet n.670645.}
\author{Rodolphe Sepulchre, Guillaume Drion, Alessio Franci}
\date{}
\maketitle

%
%
\abstract{This chapter revisits the concept of excitability, a basic system property of neurons. The focus is on excitable systems regarded as behaviors rather than dynamical systems. By this we mean open systems modulated by specific interconnection properties rather than closed systems classified by their parameter ranges. Modeling, analysis, and synthesis questions can be formulated in the classical language of circuit theory. The input-output characterization of excitability is in terms of the local sensitivity of the current-voltage relationship. It suggests the formulation of novel questions for nonlinear system theory, inspired by questions from experimental neurophysiology.}

\section{Introduction}
\label{sec:1}

In his 1996 survey paper \cite{Zames1996}, George Zames credits Charles Desoer and Mathukumalli Vidyasagar for writing the {\it ultimate text} on input-output theory of nonlinear feedback systems. This textbook was largely inspired by the engineering question of analyzing and designing nonlinear electrical circuits, a popular topic at the time. In the last decades of the century, the dominant driving application of nonlinear control theory moved from electrical circuits to robotics. The present chapter is a tribute to one of the pioneers of the input-output theory of nonlinear feedback systems. It is entirely motivated by a nonlinear electrical circuit model published in 1952 to explain the biophysical foundation of nerve excitability. The landmark paper of Hodgkin and Huxley \cite{Hodgkin1952} defined circuit theory as the modeling language of neurophysiology. Most today's questions of experimental neurophysiologists are still very naturally formulated in the language of circuit theory. But the computational push for neurophysiology {\it in silico} has progressively favored the replacement of circuit models by their state-space representations, in the form of high-dimensional models of nonlinear differential equations. The growing ease at simulating those state-space models on a personal computer is only matched by the increasing difficulty to analyse them and to resolve their inherent fragility. The difficulty of translating robustness and sensitivity questions from input-output models to state-space models is familiar to control theorists. Bridging the two worlds has been at the core of the developments of linear system theory. But progresses in the nonlinear world have been slow and limited. This bottleneck is severely restricting the possibility to analyze neuronal circuits with the tools of nonlinear state-space theory. At a broader level, this bottleneck is contributing to the gap that separates experimental neurophysiology from computational neuroscience. 

The discussion of excitability in this chapter is an attempt to revisit one of the most basic properties of biological systems in the classical language of nonlinear circuit theory. The discussion complements the presentation of excitability found in textbooks of neurodynamics. The experience of the authors in their recent work on neuromodulation \cite{Franci2014,Drion2015,Drion2015b} suggests that there is value in reopening the ultimate text of input-output nonlinear feedback systems to model excitability.  We stress the importance of {\it localized ultrasensitivity}, a defining feature of excitability that singles out a highly specific property of excitable behaviors. This property is tractable because it is amenable to local analysis. It should be acknowledged in any system theory of excitability.

\section{What is excitability?}
\label{sec:2}

Excitability is the property of a system to exhibit all-or-none response to pulse inputs. It is defined at a stable equilibrium, meaning that small inputs cause small outputs. But beyond a given threshold, the response is a large and stereotyped output. This property is primarily observed in neurons, muscle cells, and endocrine cells, where it refers to an {\it electrical} phenomenon: the input is a current, and the output is a voltage. The large stereotyped output observed in response to a current stimulus is called an action potential, or a spike. Excitability is central to physiological signaling. Ultimately, it is instrumental in converting sensory signals into motor actions. Not surprisingly, excitability is usually modeled in the language of circuit theory.

%
\begin{figure}
\includegraphics[scale=.85]{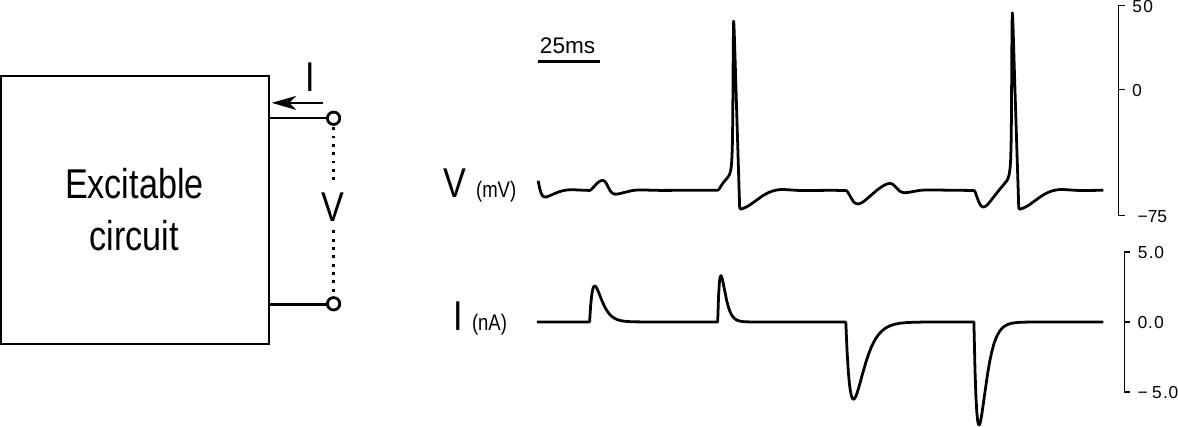}
%
%
\caption{An excitable behavior is a set of current pulses and voltage spikes defining the trajectories of a nonlinear one port circuit.}
\label{fig:1}       
\end{figure}

We regard excitability as a {\it behavioral} property in the sense of Willems \cite{Willems2007a}. An excitable behavior is the set of trajectories $(I(t),V(t))$  of a one port electrical circuit. Those trajectories are those that are observed by an electrophysiologist ; trains of pulses for the current, and trains of spikes for the voltage. A behavioral theory of excitable systems is about modeling the relationship between them with the aim of addressing questions that are system theoretic in nature:  control (what are the mechanisms that shape the behavior?), robustness (how robust is the behavior to uncertainty?), and interconnections (how to predict the behavior of the whole from the behavior of the parts?).

\begin{figure}
\includegraphics[scale=.65]{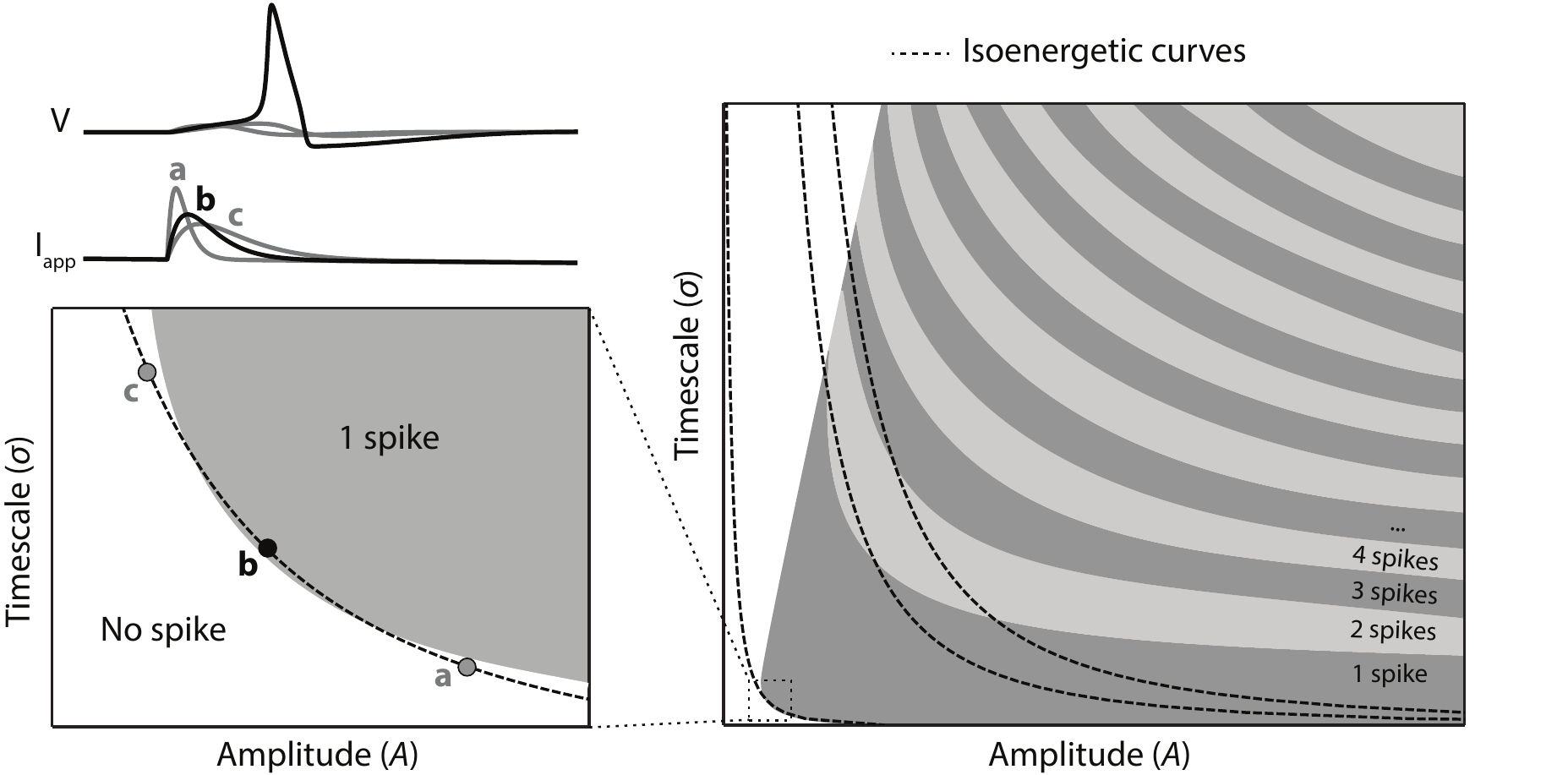}
%
%
\caption{The threshold property of an excitable circuit converts an analog pulse into a discrete number of spikes. It is localized both in amplitude and time.}
\label{fig:2}       
\end{figure}

An excitable behavior is essentially nonlinear because of the all-or-none nature of the spike. Behavioral theory is a mature theory for linear time-invariant behaviors but a theory in its infancy for nonlinear behaviors. Figure \ref{fig:2} suggests a simple way of characterizing the excitability property of a one port circuit, in analogy to the step response of a linear time invariant behavior. Here we consider a pulse current trajectory parametrized by an amplitude $A$ and a time duration $\sigma$. The figure indicates the number of spikes in the corresponding voltage trajectory.  

This representation of excitability is general and model independent. It captures the fundamental quantification property of an excitable circuit. Spikes can be regarded as discrete quantities but their number continuously depends on analog properties of the current pulses. The threshold property of an excitable system is well captured by the figure. The energy threshold is the minimum amount of charge that is necessary to trigger a spike. It endows the circuit with a characteristic {\it scale} $(A^*,\sigma^*)$, both in {\it amplitude} and in {\it time}. For a current pulse above the energy threshold, the family of pulses that can trigger a spike is {\it localized} both in {\it amplitude} and in {\it time}. For a fixed suprathreshold energy, pulses that are only localized in time or in amplitude do not trigger a spike. Energy levels are themselves quantified, meaning that an excitable circuit has a maximal spiking frequency. 
  
The spike itself is not represented in Figure \ref{fig:2} because its all-or-none nature makes it independent of the input. The input only triggers a transient excursion between an OFF state and an ON state of the circuit. The OFF state is a stable equilibrium or operating point of the circuit. The ON state is a stable limit cycle of fixed amplitude, or, less frequently, a distinct equilibrium at a significantly higher potential than the OFF state. This signature is easily identified experimentally by studying the stationary behavior of the circuit for current pulses of long duration.

There exists a large literature about the analysis of excitable models as nonlinear dynamical systems, see e.g. \cite{Izhikevich2007} and references therein. Assuming that the law of the excitable circuit is described by a nonlinear differential equation, an excitable behavior is regarded as a (closed) dynamical system by studying the trajectories of the dynamical system for a given (usually constant) current. Phase portrait analysis and bifurcation theory are the central analysis tools in this approach. Excitability is then characterized by the bifurcation that governs the transition from the OFF state to the ON state using the fixed value of the current as a bifurcation parameter. Different bifurcations define different types of excitability, associated to distinct phase portraits when modeled by second-order differential equations. While neurodynamics has been central to the development of mathematical physiology, it also suffers from limitations inherent to the dynamical systems approach. The mathematical classification is not always easy to reconcile with the neurophysiological (or behavioral) classification, and the complexity of the analysis rapidly grows with the dimension of the model. Questions pertaining to modulation, robustness, and interconnections are not easy to address in the framework of neurodynamics and call for complementary approaches.

 \section{The inverse of an excitable system}

The key advance in modeling neuronal excitability came from the voltage clamp experiment, one of the first scientific applications of the feedback amplifier. The voltage clamp experiment assigns a step trajectory to the voltage of an excitable circuit by means of a high gain feedback amplifier. The required current provides the corresponding current trajectory. In the language of system theory, the current trajectory is nothing but the step response of the inverse system.

\begin{figure}
\includegraphics[scale=.7]{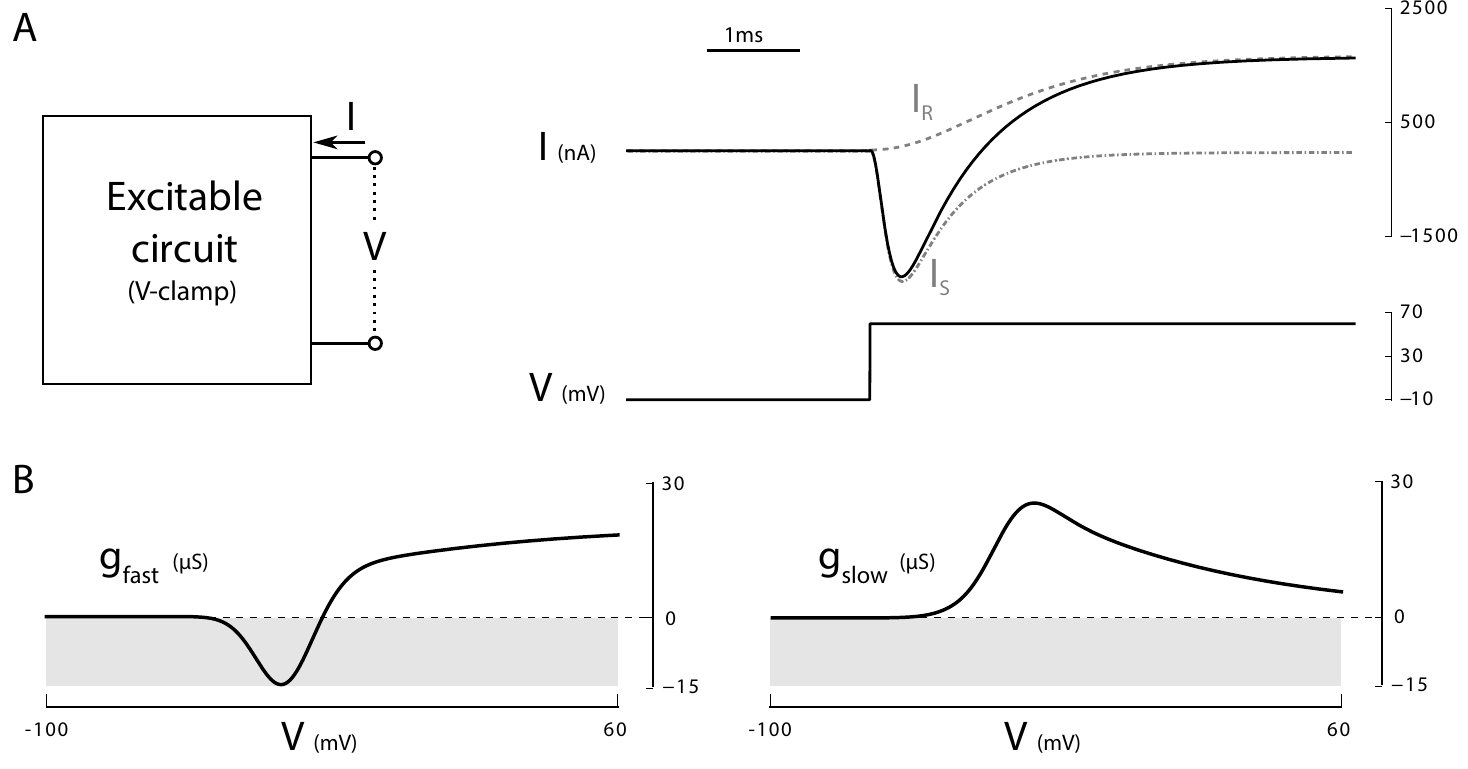}
%
%
\caption{The voltage experiment was key to modeling neuronal excitability. It provides the step response of the inverse system (\textbf{A}).  Dynamic input conductances (DICs)  extract key properties of the inverse system response as a function of voltage. (\textbf{B}). The trajectory and DICs shown are computed from the Hodgkin Huxley model.}
\label{fig:3a}       
\end{figure}

The step response in Figure \ref{fig:3a}A is typical of a transfer function with a fast right-half plane zero and a slow left-half plane pole: the sign of the low frequency (or static) gain is opposite to the sign of the high-frequency (or instantaneous) gain. This property led Hodgkin and Huxley to identify the distinct roles of a slow and of a fast currents ($I_{early}$ and $I_{late} $ in the terminology of \cite{Hodgkin1952b}) as a key mechanism of excitability. The fast right-half plane zero of the voltage-driven circuit corresponds to a fast unstable pole of the current-driven excitable system, whereas the slow left-half plane  pole corresponds to a slow left-half plane zero. Hodgkin and Huxley also observed that this essential feature of the step response is voltage-dependent. It holds for a step voltage around the resting potential, but it disappears if the step voltage is repeated around a higher potential. At higher values of the potential, the step response becomes the stable response of a slow first-order system (not shown here, but abundantly illustrated in \cite{Hodgkin1952b}). In other words, the non-minimum phase nature of the step response shown in Figure \ref{fig:3a} only holds in a narrow voltage range.

To date, the voltage clamp experiment remains the fundamental experiment by which a neurophysiologist studies the effect of neuromodulators or the role of particular ion channels in a specific neuron. The recent paper \cite{Drion2015} by the authors proposes that modulation and robustness properties can indeed be studied efficiently via the dynamic conductances of the neuron. Dynamic conductances extract from small step voltage clamp trajectories the {\it quasi-static} conductance $\frac{\Delta I}{\Delta V}$ in different time scales. Figure~\ref{fig:3a}B shows the {\it fast} ($g_{fast}$) and {\it slow} ($g_{slow}$) dynamic conductances of Hodgkin and Huxley model. The non-minimum phase voltage step response translates into a voltage range where the fast (or instantaneous) conductance is negative, whereas the slow conductance is positive. The dynamic input conductance curves quantify the temporal and voltage dependence of the conductances. The fast dynamic conductance is negative close to the resting potential, whereas the slow dynamic conductance is positive everywhere. Those features are the essential signature of an excitable circuit. In particular, the zero crossing of the fast conductance is an excellent predictor of the threshold voltage and the fundamental signature of the {\it localized sensitivity} of the circuit. A small conductance means ultrasensitivity of the circuit with respect to current variations. The voltage clamp experiment identifies the temporal and amplitude window of ultrasensitivity of an excitable circuit.

 \section{A circuit representation and a balance of positive and negative conductances}

An excitable circuit is made of three distinct elements: a passive circuit, a switch, and a regulator. Each element is itself a one port circuit and the three elements are interconnected according to Kirchoff law. 

The {\it passive circuit} accounts for the passive properties of the excitable behavior. Its static behavior is strictly resistive, hence the monotone current-voltage (I-V) relationship. Its dynamic behavior is strictly passive, meaning that the circuit can only dissipate energy. In the language of dissipativity theory, the change in internal energy stored in the capacitor is upper-bounded by the external supplied power \cite{Willems1972}. 

The {\it switch} accounts for the large voltage transient of the spike. Its static behavior is characterized by a range of negative conductance. It is the destabilizing element of the excitable circuit and it requires an active source. The activation is however {\it localized}, meaning that the negative conductance of the switch can overcome the positive conductance of the passive circuit only within a local amplitude and temporal range.  

The {\it regulator} accounts for the repolarization of the circuit following a spike. In particular, it ensures a refractory period following the spike, which contributes to the {\it all-or-none} nature of the spike and to its temporal scale: two consecutive spikes are always separated by a minimal time interval. The regulatory element is a distinct source of dissipation, that continuously modulates the balance between the positive conductance of the passive circuit and the negative conductance of the switch.

  \begin{figure}
\includegraphics[scale=.85]{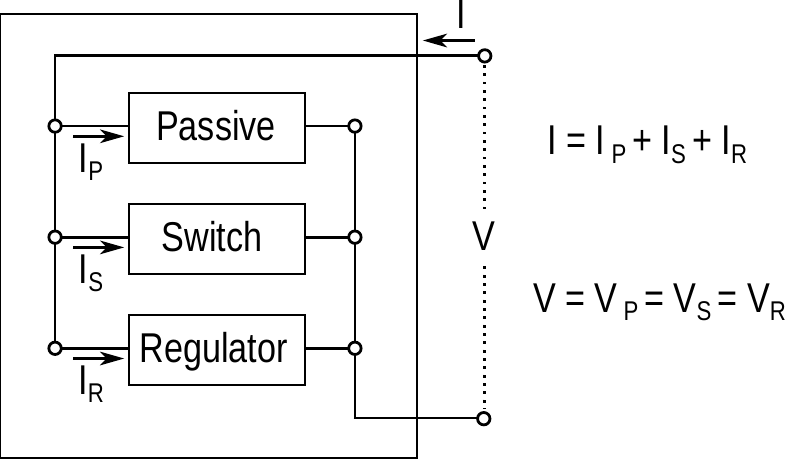}
%
%
\caption{The three circuit elements of an excitable one-port circuit.}
\label{fig:3}       
\end{figure}

The static model of an excitable circuit is a nonlinear resistor, characterized by its so-called $I-V$ curve. This curve can be either monotonically increasing, or hysteretic if the negative resistance of the switch locally overcomes the positive resistance in static conditions. Hysteresis of the I-V curve is not necessary for excitability \cite{Drion2015b}, a clear evidence that excitability is a {\it dynamical} phenomenon. In classical circuit theory, the dynamics of the circuit is captured by a small-signal analysis around operating points. The {\it admittance} of the circuit at a given operating point is the dynamic generalization of its conductance. It is the frequency response of the linearized behavior $\delta I = G( j \omega) \delta V$ around a given operating point $(I,V)$. The admittance is a complex number that depends both on the amplitude $V$ and of the frequency $\omega$. 

The admittance of a passive circuit is positive real, that is, its real part is nonnegative at all frequencies. The regulatory element preserves this property if it is itself passive. In contrast, the switch element creates an amplitude and frequency range where the real part of the admittance becomes negative. It is the only destabilizing element of the circuit. A clear local signature of excitability is therefore a localized amplitude and frequency range where the real part of the admittance becomes negative.  

The characterization of an excitable circuit from its admittance properties is not limited to low-dimensional models amenable to phase portrait analysis. The dynamic input conductances discussed in Section \ref{sec:2} are snapshots of the admittance in different time scales.

%
%
%

\section{A mixed feedback motif and a robust balance property}
\label{sec:4}

Due to its negative conductance, the switch of an excitable circuit acts as a source of positive feedback. Due to its positive conductance, the regulator of an excitable circuit acts a source of negative feedback.  As a consequence, an excitable circuit always admits the representation of a passive system surrounded by two distinct feedback loops of opposite sign. This representation is important because it coincides with the excitatory-inhibitory (E-I) feedback motif found in many biological models. The excitatory feedback loop often models an autocatalytic process whereas the inhibitory feedback loop often corresponds to a regulatory process.

\begin{figure}
\centering
\includegraphics[width=0.88\linewidth]{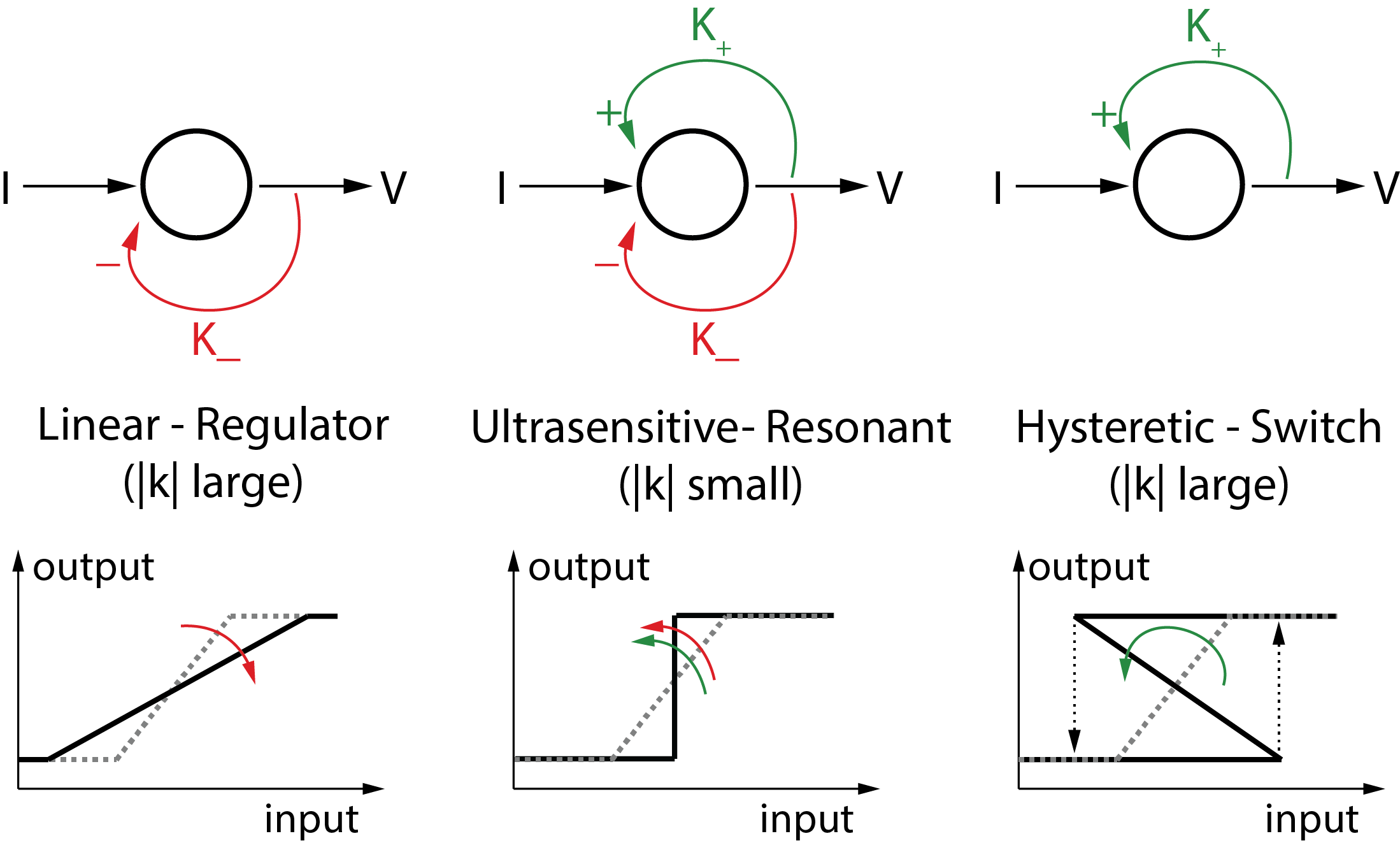}
\caption{\textbf{Regulating the balance between positive and negative feedback can switch a system between linear, ultrasensitive and hysteretic states.} Top, sketches of the systems composed of a negative feedback (left), a positive feedback (right), or both (center). Bottom, input/output relationships in the three cases. The dashed grey lines show the open-loop relationships, and the full black lines the closed-loop relationships.} \label{fig:4}
\end{figure}

The balance between positive and negative feedback is key to the localized sensitivity of an excitable circuit. The static picture is the one of the mixed feedback amplifier illustrated in Figure~\ref{fig:4}. When negative feedback dominates, the circuit is purely resistive and the behavior is analog. More negative  feedback enlarges the linearity range of the circuit and decreases its input-output sensitivity. In contrast, when positive feedback dominates, the circuit is hysteretic and the behavior is quantized. More positive feedback enlarges the hysteretic range and decreases its input-output sensitivity in the OFF and ON mode. When positive and negative feedback balance each other, the circuit becomes characterised by a tiny range of ultrasensitivity. 

An excitable circuit offers a versatile architecture to tune ultrasensitivity by balancing positive and negative feedback. The switch ensures a local range in time and amplitude where the circuit behaves as a hysteretic switch. The regulator ensures that, away from a local range, the circuit behavior is resistive and linear. By continuity of the feedback gain, the circuit must be ultrasensitive along trajectories that connect the switch-like and the linear-like behaviors. Suprathreshold current pulses and the corresponding spikes are examples of such trajectories.

The feedback representation of an excitable circuit highlights the robustness of achieving ultrasensitivity by a balance of feedback. Ultrasensitivity must exist provided that there exists a local range in amplitude and time where positive feedback dominates negative feedback. For the admittance of the circuit along its I-V curve, this means that there must exist a voltage range and a frequency range where the negative real part of the switch admittance exceeds the positive real part of the passive admittance. For the supplied energy, this means that there must exist trajectories in a local amplitude and temporal range where the overall circuit is active, that is, the energy supplied by the switch exceeds the energy absorbed by the passive admittance.

Time scale separation between a fast switch and a slow regulator enhances the robustness of excitability, creating a two time-scale circuit that behaves as a bistable switch in the fast time scale and as a linear resistive cicruit in the slow time scale. In the spike of Figure \ref{fig:1}, the fast behavior is the upstroke, whereas the repolarization is the slow behavior. As the time-scale separation between the switch and the regulator decrease, the distinction between the "switch-like" and "linear-like" parts of an excitable behavior become progressively blurred. The localization of excitability requires a hierarchy between the positive and the negative feedback: the range where the positive feedback gain exceeds the negative must be narrow relative to the negative feedback range. The feedback motif of an excitable circuit is thus {\it fast and localized positive} feedback balanced by {\it slow and global negative} feedback.

\section{Models of excitability}
\label{sec:5}

\subsection{FitzHugh-Nagumo circuit}

%

Figure \ref{fig:5}.A  shows an elementary circuit fitting the requirements of Figure \ref{fig:3}: the parallel connection of a capacitor (the passive element), a static diode with a cubic  $I-V$ characteristic (the switch), and an RC  branch (the regulator). This circuit was first studied by Nagumo et al. \cite{Nagumo1962}, following the proposal of FitzHugh \cite{FitzHugh1961} to study excitability through a minor modification of Van der Pol oscillator. The motivation in both papers was to extract a simple qualitative model of Hodgkin-Huxley model. FitzHugh-Nagumo model admits the state-space representation
\begin{equation}
\label{FNcircuit}
\begin{array}{rcl}
C \dot V& = & - i_s - i_r+  I   \\
L \dot i_r & = & -R i_r + V, \\
i_s & = & \frac{V^3}{3}-kV
 \end{array}
\end{equation}
Its phase portrait has been a key paradigm to explain excitability ever since. See for instance \cite{Izhikevich2006} and references therein.
 
In the configuration where the static conductance of the regulator exceeds the negative conductance of the diode, i.e. $k < \frac{1}{R}$, the static I-V curve is monotone. The circuit is excitable when the capacitance $C$ is small relative to the time constant $\tau=\frac{L}{R}$ of the regulatory element. The circuit can then be analyzed as a fast-slow system. The fast subsystem is made of the capacitor and the switch element. Its static I-V curve is the cubic characteristic of the diode. This circuit is a simple bistable device. The slow subsystem is made of the regulatory inductive element, which is a linear first-order lag. The fast-slow behavior is ultrasensitive in the amplitude and voltage range where the real part of the admittance
$$
G(j\omega;\bar V)= C j \omega + (\bar V^2-k) + \frac{1}{L j\omega + R}
$$
is close to zero. Sensitive trajectories include fast current pulses applied near the local extrema of the cubic characteristic of the switch. 

\subsection{Hodgkin-Huxley circuit}

\begin{figure}
\includegraphics[width=0.8\textwidth]{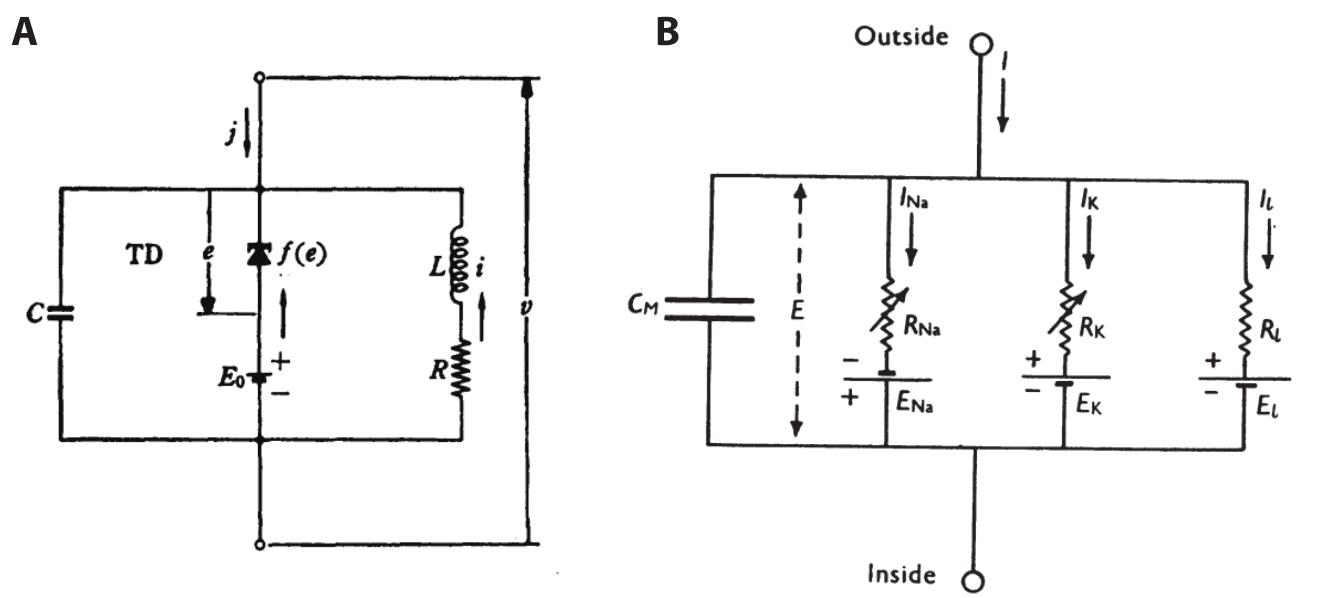}
%
%
\caption{A. Fitzthugh-Nagumo circuit (from  \cite{Nagumo1962}). B. Hodgkin-Huxley circuit  (from \cite{Hodgkin1952}).}
\label{fig:5}       
\end{figure}

Figure \ref{fig:5}.B reproduces the first figure from the landmark paper of Nobel prize winners Hodgkin and Huxley \cite{Hodgkin1952}. The circuit models the excitability of a neuron. The passive element is the RC circuit made of the capacitor (modeling the cellular membrane) and the leaky current $I_L$.  The switch element is the sodium current $I_{Na}$. The regulatory element is the potassium current $I_K$. Using the voltage clamp experiment, the authors identified the following model for the two ionic currents:
\begin{eqnarray*}
 I_K = \bar g_K n^4 (V-V_K) \\
\tau_n (V) \dot n = -n + n_\infty(V)
\end{eqnarray*}
and
\begin{eqnarray*} I_{Na} = \bar g_{Na} m^3 h (V-V_{Na}) \\
\tau_m (V) \dot m = -m+ m_\infty(V) \\
\tau_h (V) \dot h = -h + h_\infty(V)
\end{eqnarray*}

\begin{figure*}
\centering
 \includegraphics[width=0.95\linewidth]{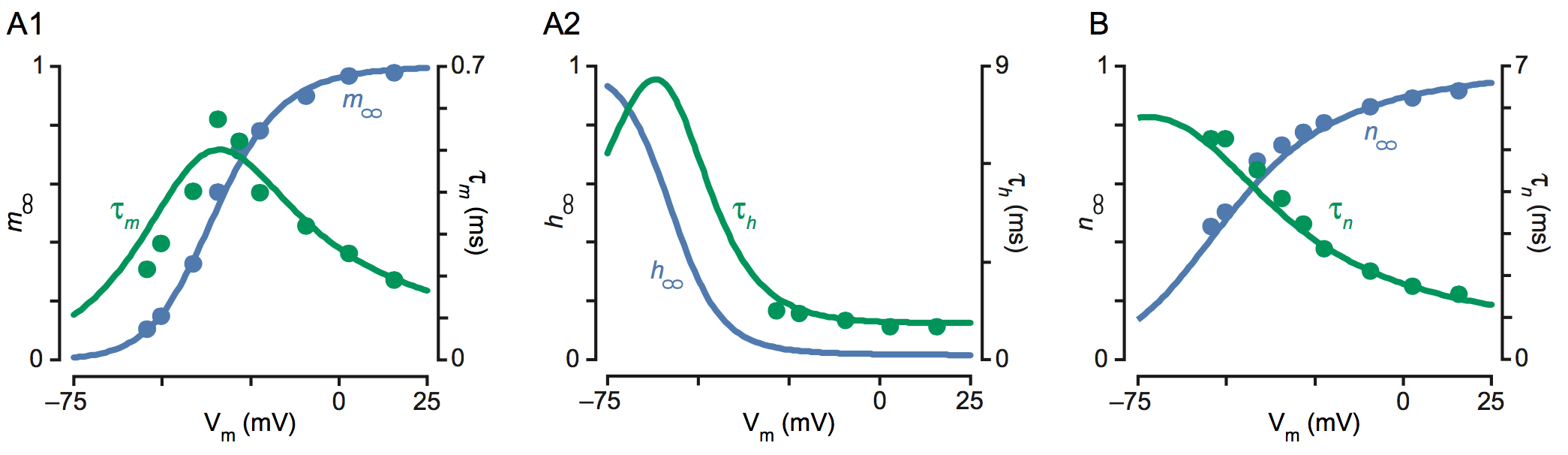}
\caption{\textbf{Voltage-dependence of the time-constants and the static gains of the Hodgkin-Huxley model \cite{Hodgkin1952}}. The blue curves correspond to the sodium steady-state activation $m_{\infty}(V_m)$, the sodium steady-state inactivation $h_{\infty}(V_m)$ and the potassium steady-state activation $n_{\infty}(V_m)$. The green curves correspond to the sodium activation time-constant $\tau_m(V_m)$, the sodium inactivation time-constant $\tau_h(V_m)$ and the potassium activation time-constant $\tau_n(V_m)$. The dots represent corresponding experimental data. Reproduced from \cite{Byrne2014}.} \label{fig:7}
\end{figure*}

The state variables $m$, $n$, and $h$ are {\it gating variables} in the range $[0,1]$ introduced to model the amplitude and temporal dependence of the ionic conductances. The voltage dependent time constants and gains of the gating variables were obtained by curve fitting, see Figure \ref{fig:7}. The admittance of the potassium current is
$$ g_K(\bar V; j \omega) = \bar g_K (\bar V- V_K)  4 n^3(\bar V) \frac{n_\infty'(\bar V)}{1 + \tau_n(\bar V) j \omega} $$
It is positive real provided that $\bar V \ge V_K$, that is, whenever the potassium current is an outward current, which is always the case in physiological conditions.
The admittance of the sodium current is
$$ g_{Na}(\bar V; j \omega) = \bar g_{Na} m^2(\bar V) (\bar V- V_{Na})  ( 3  h(\bar V)  \frac{m_\infty'(\bar V)}{1 + \tau_m(\bar V) j \omega}  +  m(\bar V)  \frac{h_\infty'(\bar V)}{1 + \tau_h(\bar V) j \omega} )
$$
At any voltage and any frequency, it is the sum of two terms of opposite real part. Whenever the sodium current is an inward current, i.e. $\bar V \le V_{Na}$, which is always the case in physiological conditions, the first term is negative real whereas the second term is positive real. Looking at Figure \ref{fig:7}, it is obvious that the negative real term largely dominates the positive real term in a voltage range that includes the resting potential (around $-70mV$ and  the high frequency range $\approx 1 ms^{-1}$). The sodium current thus acts as a negative resistance switch in the fast dynamic range of the activation variable $m$, whose time constant is about ten times smaller than the other gating  time constants near the resting potential. In turn, the sodium {\it inactivation} variable $h$ and the potassium {\it activation} variables both contribute to the negative feedback that regulates the refractory period of the spike. 

It is important to observe that the balance of positive and negative feedback responsible for the ultrasensitivity of the circuit is robust to the details of the modeling. It rests entirely on the {\it localization} of the positive feedback in a specific voltage range ({\it near} the resting potential) and in a specific frequency range (about one decade above the cutoff frequency of the regulatory elements). This localization makes the excitability robust, by ensuring a range of ultrasensitivity (i.e. balance between positive and negative feedback) independent of the modeling details of the circuit.

FitzHugh-Nagumo circuit captures the excitability of Hodgkin-Huxley circuit by modeling the sodium activation as an instantaneous negative resistance diode and the sodium inactivation and potassium activation as a slow linear regulatory feedback. The reader will notice that this simplification introduces two artifacts. First, the time scale of the positive feedback must be fast relative to the time scale of negative feedback but should not be constrained to be instantaneous. In fact, this time scale is a critical feature of excitability as it sets the temporal localized window of excitability. Second, the time-scale separation between slow and fast gating variables in Hodgkin-Huxley circuit is only observed in a narrow voltage range around the resting potential. It vanishes at higher voltages, which means that the ultrasensitivity region is confined to the resting potential voltage range. There is no ultrasensitivity during the spike. In contrast, because the voltage dependence of time constants is ignored in FitzHugh-Nagumo circuit, the spikes have a range of ultrasensitivity both
in the subthreshold and suprathreshold voltage ranges.
 
%
%
%
%
%

\section{Conclusion}
 \label{sec:6}
 
%
%
%
%
%
%

Excitability is a behavior at the core of biology. The presentation in this chapter emphasizes that the core property of an excitable circuit is a localized ultrasensitivity of the current-voltage relationship: small current variations are largely amplified in a specific temporal and amplitude range. This property can be quantified by the elementary concept of dynamic input conductance, which is nothing but the local gain of the inverse system computed at a given voltage and in a given time scale. Excitable circuits can be modulated by shaping their dynamic conductance, that is, by localizing the windows of low conductance (i.e. high sensitivity). Excitable circuits can be interconnected to create behaviors with localized and overlapping windows of ultrasensitivity. The recent study by the authors of modulation and robustness of bursting \cite{Franci2014} is a first step in that direction.

\bibliographystyle{unsrt}
\bibliography{book_chapter}
\end{document}